\documentclass[12pt,thmsa]{article}
\usepackage{sw20aip}
\usepackage{amssymb}
\usepackage{graphicx}
\usepackage{layout}

\begin{document}

\title{Testing gravitomagnetism on the Earth}
\author{A. Tartaglia, M. L. Ruggiero \\
Dip. Fisica, Politecnico, and INFN, Torino, Italy\\ e-mail:
tartaglia@polito.it ; ruggierom@polito.it}

\begin{abstract}
The paper contains a proposed experiment for testing the
gravitomagnetic effect on the propagation of light around a
rotating mass. The idea is to use a rotating spherical
laboratory-scale shell, around which two mutually orthogonal
lightguides are wound acting as the arms of an interferometer.
Numerical estimates show that time of flight differences between
the equatorial and polar guides could be in the order of $\sim
10^{-20}$ s, actually detectable with sensitivity perfectly
comparable with those expected in gravitational wave detection
experiments. \end{abstract}

\maketitle

\section{Introduction}

The gravitomagnetism is the part of the gravitational field which
displays a solenoidal character similar to that of the magnetic
field \cite{gravimag}. It is embedded in general relativity,
however its effects are usually much less relevant than those of
the gravitoelectric (radial) part of the field. Gravitomagnetism
is expected to influence the precession of orbiting gyroscopes
(Lense-Thirring effect \cite{lense}), the synchronization of
clocks in the field of rotating masses \cite{mashhoon},
\cite{tartaglia}, the time of flight of light around spinning
bodies \cite{tartaglia2}.

Actually in the field of the Earth the relevance of
gravitomagnetic effects is extremely small and the only attempts
to detect them have since been limited to the precession of the
nodes of the orbits of LAGEOS\ satellites \cite{ciufolini} and to
the precession of gyroscopes carried by the space shuttle
\cite{gpb}.

Here we propose a ground based experiment exploiting the effect on
the time of flight of light rays, induced by a rotating mass.
Actually, as we shall see, the time of flight is influenced both
by the very mass $M$ of the central body and by its angular
momentum density, expressed by the parameter $a=J/(Mc)$ i.e. times
the ratio between the angular momentum and the product of the mass
by the speed of light. However when considering the time of flight
difference between an equatorial and a polar circular trajectory,
it turns out, at the lowest significant order, to be proportional
to $a^{2}$.

The actual value of $a$ depends on the geometry of the source and
its angular velocity. The highest results are obtained in thin
spherical shells. For the whole Earth $a$ is in the order of 4 m;
at the laboratory scale it is some orders of magnitude lower,
however we shall show that the final value for the time difference
is within the sensitivity range of interferometric techniques
presently available and under consideration for gravitational
waves detectors. The expected cost for the proposed on Earth
experiment should in turn be much lower than those in space.

\section{The time difference}

In an axially symmetric stationary gravitational field the null
interval is written:
\[
0=g_{tt}dt^{2}+2g_{t\phi }dtd\phi +g_{rr}dr^{2}+g_{\theta \theta
}d\theta ^{2}+g_{\phi \phi }d\phi ^{2}
\]
where the metric elements do not contain either $t$ or $\phi $.
Considering a circular path the (coordinate) time of flight for an
equatorial revolution ($\theta =\pi /2$) is \cite{tartaglia3}

\begin{equation}
T_{e}=2\pi \frac{\mp g_{t\phi }+\sqrt{g_{t\phi }^{2}-g_{tt}g_{\phi \phi }}}{%
g_{tt}}  \label{equatoriale}
\end{equation}
The $-$ sign stands for co-rotation, $+$ means counter-rotation.

Using Boyer-Lundquist coordinates in a Kerr metric
(\ref{equatoriale}) becomes
\begin{equation}
T_{e}=\frac{2\pi }{c^{2}}\frac{\mp \frac{2GMa}{cr}+c\sqrt{\left( \frac{2GMa}{%
c^{2}r}\right) ^{2}+\left( 1-\frac{2GM}{c^{2}r}\right) \left( r^{2}+a^{2}+%
\frac{2GMa^{2}}{c^{2}r}\right) }}{1-\frac{2GM}{c^{2}r}}
\label{kerr}
\end{equation}

The other configuration we are considering is a fixed azimuth
polar circular trajectory. Now it is
\[
T_{p}=\int_{o}^{2\pi }\sqrt{-\frac{g_{\theta \theta
}}{g_{tt}}}d\theta
\]
or explicitly
\begin{equation}
T_{p}=\frac{1}{c}\int_{o}^{2\pi }\sqrt{\frac{r^{2}+a^{2}\cos ^{2}\theta }{1-%
\frac{2\frac{GM}{c^{2}}r}{r^{2}+a^{2}\cos ^{2}\theta }}}d\theta
\label{polarkerr}
\end{equation}

In a weak field and introducing the small parameters $\mu
=\frac{2GM}{c^{2}r} $ and $\alpha =a/r$ (\ref{kerr}) and
(\ref{polarkerr}) become
\begin{eqnarray*}
T_{e} &=&\frac{2\pi }{c}R\left( 1+\frac{1}{2}\alpha ^{2}+\allowbreak \frac{1%
}{2}\mu \right) \\ T_{p} &=&\frac{R}{c}\int_{0}^{2\pi }\left(
\allowbreak 1+\frac{1}{2}\alpha ^{2}\cos ^{2}\theta
+\frac{1}{2}\mu \right) d\theta \\ &=&\allowbreak 2\pi
\frac{R}{c}\left( 1+\frac{1}{2}\mu +\frac{1}{4}\alpha ^{2}\right)
\end{eqnarray*}
Finally we obtain
\begin{equation}
\Delta T=T_{e}-T_{p}=\allowbreak \frac{1}{2}\frac{\pi }{cR}a^{2}
\label{differenza}
\end{equation}
which, as can be seen, depends only on $a^{2}$ (the terms
containing the mass mix it with $a$ and are smaller).

\section{Laboratory scale}

For an homogeneous steadily rotating sphere it is
\begin{equation}
a_{f}=\frac{2}{5}\frac{R^{2}}{c}\Omega  \label{pianeti}
\end{equation}
where $\Omega $ is the angular speed of the sphere.

The value of $a$ can be increased a bit considering instead of a
sphere a hollow spherical thin shell. In that case one has
\begin{equation}
a_{h}=\frac{2}{3}\frac{R^{2}}{c}\Omega  \label{guscio}
\end{equation}
Now however the mass, for the same external radius, is much lower
than before. Actually
\[
\frac{M_{h}}{M_{f}}=\allowbreak 3\frac{h}{R}
\]
where $h$ is the thickness of the shell ($h<<R$).

Applying (\ref{pianeti}) to the Earth the result is
\begin{equation}
a_{E}=3.9\ m \nonumber
\end{equation}

Other examples are Jupiter or the Sun \cite{tartcqg17}:
\begin{eqnarray*}
a_{J} &=&1.2\times 10^{3}\ m \\ a_{S} &=&3.0\times 10^{3}\ m
\end{eqnarray*}

In the laboratory one can of course expect much lower values. Let
us consider a hollow sphere as the source of the effect. The $a$
value is limited in practice by the strength of the wall of the
shell. In fact the resulting centrifugal force on a hemisphere is
\[
F_{c}\allowbreak =\pi \rho h\Omega ^{2}R^{3}
\]
($\rho $ is the density of the material).

The corresponding average tension induced in the wall of the shell
is
\[
<\sigma >=\frac{\pi \rho h\Omega ^{2}R^{3}}{2\pi Rh}=\allowbreak \frac{1}{2}%
\rho \Omega ^{2}R^{2}
\]
The maximum stress is attained at the equator, being in the order
(unidimensional stresses are assumed):
\[
\sigma _{m}=\allowbreak \rho \Omega ^{2}R^{2}
\]

If $\sigma _{m}$ coincides with the allowable resistance of the
material the maximal peripheral velocity is $v_{m}=\sqrt{\sigma
_{m}/\rho }$. The attainable value of $a_{h}$ can consequently be
written in terms of the properties of the material:
$a_{h}=\frac{2}{3}\frac{R}{c}\sqrt{\sigma _{m}/\rho }$.

Finally the time difference (\ref{differenza}) becomes
\begin{equation}
\Delta T=\allowbreak \frac{2}{9}\frac{\pi }{c^{3}}\frac{R^{2}}{R_{l}}\frac{%
\sigma _{m}}{\rho }  \label{pratico}
\end{equation}
Here $R_{l}$ is the radius of the light's path (a little bit greater than $R$%
).

Considering composite materials $\sigma _{m}$ can be as high as
2000 MPa, with a density $\rho \sim 1700$ kg/m$^{3}$ \cite{romeo}.
These values lead to (assuming, just to fix ideas, $R=1$ m)
\begin{eqnarray*}
a_{h} &=&2.4\times 10^{-6}\ m \\ \Delta T &=&3\times 10^{-20}\ s
\end{eqnarray*}
Using visible light the relative phase shift corresponding to the
time of flight difference is in the order of $10^{-5}$.

\section{Proposing an actual experiment}

A phase shift like the one computed in the previous section is of
the same order of magnitude as the phase differences expected in
gravitational waves interferometric detection experiments
\cite{virgo}. The advantage now could be the comparatively small
size of the apparatus. The idea is to have as a gravitomagnetic
source a spinning thin spherical shell; to fix a hypothesis it
could have a $1$ m radius and a wall thickness of $1$ mm or less.
Such object would weigh not more than 209 N and should rotate at a
maximum angular speed of $\Omega _{m}\simeq 10^{3}$ rad/s. Two
circular non rotating light guides should contour the sphere, one
at the equator, the other through the poles, as in figure 1. A
laser beam would be split at $A$, the two resulting secondary
beams would be guided along the two circular paths and finally
would be led to interfere at $B$.

\begin{figure}[top]
\begin{center}
\includegraphics[width=9cm,height=7cm]{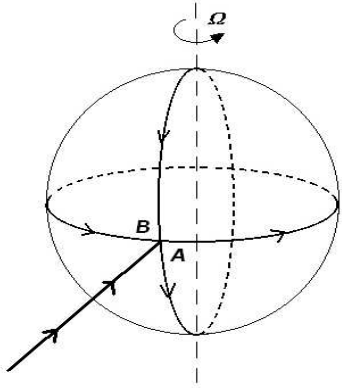}
\caption{ The hollow sphere rotates at the angular speed $\Omega $%
. The two circular wave guides are fixed. A primary light beam is split in $%
A $ to follow separately the equatorial and the polar
trajectories, then the beam is recombined interfering in $B.$}
\label{fig:fig1}
\end{center}
\end{figure}

The beam intensity relative change at the interference $\delta
I/I$ is related to the phase shift $\delta \Phi $ according to
\[
\frac{\delta I}{I}=\frac{1}{2}\left( 1-\cos \delta \Phi \right)
\]
Using the estimates of the previous section, this means
\[
\frac{\delta I}{I}\sim 10^{-9}
\]

It is not possible to extract such an intensity fluctuation if it
is static. This means that we need modulating it in time; this
result can be achieved periodically varying the angular speed of
the sphere. In a sense we would be simulating the effect of a
gravitational wave of very low frequency (reasonably fractions of
a Hz).

\section{Conclusion}

We have shown that it is possible to realize a ground based
experiment to reveal gravitomagnetic effects using a laboratory
size rotating mass. In fact available materials (composite carbon
fibers high resistance materials) and available technologies for
detection of very small periodically varying intensity
perturbations in a light signal do allow for the possibility to
measure the time difference (\ref{differenza}) and consequently
the influence of the angular momentum density around a spinning
body. Of course a lot of technical details need being clarified,
but with no higher difficulty than the problems implied by
interferometric detection of gravitational waves. The advantage of
our proposal would be to have a cost presumably much lower than
other experiments requiring satellites and space missions,
revealing a small scale weak general relativistic effect.


\begin{thebibliography}{99}
\bibitem{gravimag}  B. Mashhoon, F. Gronwald, H.I.M. Lichtenegger, submitted
to Proc. Bad Honnef Meeting on: GYROS, CLOCKS, AND
INTERFEROMETERS: TESTING GENERAL RELATIVITY IN SPACE (22 - 27
August 1999; Bad Honnef, Germany) and Los Alamos Archives
\textit{gr-qc/9912027; }J.M. Cohen and B. Mashhoon, \textit{Phys.
Lett. A} \textbf{181}, 353 (1993).

\bibitem{lense}  H. Thirring, \textit{Phys. Z.} \textbf{19}, 33 (1918);
\textbf{22, }29\textbf{\ }(1921); J. Lense and H. Thirring, \textit{Phys. Z}%
., \textbf{19}, 156 (1918); B. Mashhoon, F.W. Hehl and D.S. Theiss, \textit{%
Gen. Rel. Grav.} \textbf{16}, 711, (1984).

\bibitem{mashhoon}  B. Mashhoon, F. Gronwald,F.W.Hehl, D. S. Theiss, \textit{%
Ann. Phys}. \textbf{8, }135 (1999) and Los Alamos Archives \textit{%
gr-qc/9804008}

\bibitem{tartaglia}  A. Tartaglia, \textit{Phys.Rev. D }\textbf{58, }064009
(1998) and Los Alamos Archives \textit{gr-qc/9806019}

\bibitem{tartaglia2}  A. Tartaglia, \textit{Class. Quantum Grav}. \textbf{%
17, }783 (2000) and Los Alamos Archives\textit{\ gr-qc/9909006}

\bibitem{ciufolini}  I. Ciufolini, \textit{Class. Quantum Grav.}, \textbf{17}%
, 2369 (2000)

\bibitem{gpb}  S. Buchman \textit{et al.}, \textit{Advances in Space Research%
}, \textbf{25} Issue 6, 1177 (2000)

\bibitem{tartaglia3}  A. Tartaglia \textit{Class. Quantum Grav.} \textbf{17,
}2381 (2000)

\bibitem{tartcqg17}  A. Tartaglia, Class. Quantum Grav. \textbf{17, }783
(2000)

\bibitem{romeo}  B. Fornari , D. Dosio , G. Romeo : ''Characterization of a
State of the Art UHM CFRP System for Satellite Application''.
Proc. of Int. Symposium on Advanced Materials for Lightweight
Structures '94. ESTEC, Noordwijk (NL), March 1994. ESA-WPP-070,
1994, pp. 569-575. Noordwijk (NL), 1994.

\bibitem{virgo}  R. Passaquieti \textit{et al., Nuclear Physics B} - \textit{%
Proceedings Supplements}, \textbf{85} Issues 1-3, 241 (2000)

\bibitem{interf}  P.R. Saulson, \textit{Class. Quantum Grav. }\textbf{17, }%
2441 (2000)

\bibitem{interf1}  P. Fritschel \textit{et al., Phys. Rev. Lett. }\textbf{%
80, }3181 (1998)
\end{thebibliography}
\end{document}